\documentclass{webofc}
\usepackage[varg]{txfonts}   % Web of Conferences font
\usepackage{listings}
\usepackage{pythonhighlight}
\begin{document}
\title{RootInteractive tool for multidimensional statistical analysis, machine learning and analytical model validation}

\author{\firstname{Marian} \lastname{Ivanov}\inst{1}\fnsep\thanks{\email{marian.ivanov@cern.ch}} \and
        \firstname{Marian} \lastname{Ivanov}\inst{2}\fnsep\thanks{\email{marian.i@cern.ch}} 
\and
        \firstname{Giulio} \lastname{Eulisse}\inst{2}\fnsep\thanks{\email{Giulio.Eulisse@cern.ch}}       
}

\institute{GSI Darmstadt\and
           UK Bratislava\and
           CERN
          }

\abstract{%
The ALICE experiment \cite{aamodt2008alice} at CERN's LHC is specifically designed for investigating heavy ion collisions. The upgraded ALICE accommodates a tenfold increase in Pb–Pb luminosity and a two-order-of-magnitude surge in minimum bias events. To address the challenges of high detector occupancy and event pile-ups, advanced multidimensional data analysis techniques, including machine learning (ML), are indispensable. Despite ML's popularity, the complexity of its models presents interpretation challenges, and oversimplification in analysis often leads to inaccuracies.

Our objective was to develop RootInteractive, a tool for multidimensional statistical analysis. This tool simplifies data analysis across dimensions, visualizes functions with uncertainties, and validates assumptions and approximations. In RootInteractive, it is crucial to easily define the functional composition of analytical parametric and non-parametric functions, exploit symmetries, and define multidimensional "invariant" functions and corresponding alarms.

RootInteractive \cite{RootInteractive} adopts a declarative programming paradigm, ensuring user-friendliness for experts, students, and educators. It facilitates interactive visualization, n-dimensional histogramming/projection, and information extraction on both Python/C++ server and Javascript client. The tool supports client/server applications in Jupyter or standalone client-side applications. Through data compression, datasets with O($10^7$) entries and O(25) attributes can be interactively analyzed in a browser with O(0.500-1 GB) size. Representative downsampling and reweighting/pre-aggregation enable the effective analysis of one year of ALICE data for various purposes.

}
\maketitle
\section{Introduction}
\label{intro}
RootInteractive, developed within the ALICE collaboration, is a tool for advanced multidimensional interactive statistical analysis. It effectively addresses the challenges of the high interaction rates of ALICE data taking during LHC Run 3. These challenges include for example event pile-up, space point distortions in the Time Projection Chamber (TPC) detector due to the accumulated space charge, electronic baseline fluctuations in the TPC, and other distortions.
A deep understanding of the detector system, MC simulations and calibration performance is essential for effective use of machine learning in physics analysis. Our goal is to provide a solution that simplifies multidimensional data analysis:

\begin{itemize}
    \item Fitting and visualizing N-dimensional functions taking into account uncertainties and biases.
    \item Streamline validation of assumptions, numerical evaluations and differential model comparisons, enabling function composition for different mathematical functions and error propagation.
    \item Rapid feedback, reducing analysis time from weeks to seconds for interactive expert discussions.
    \item Support for multi-dimensional parametric optimization.
    \item User-friendly configuration options for visualisation of unbinned and binned data, interactive multidimensional histograms and projections. It also allows the derivation of aggregated information accessible on both server (Python/C) and client (JavaScript) platforms.
    \item To facilitate the creation of stand-alone client-side applications (HTML documents) without the need to install additional software.
\end{itemize}

RootInteractive's core philosophy, encapsulated in the motto "seeing is believing", emphasizes the importance of queries, iterative interactions, and differential comparisons in understanding complex data. Within ALICE, RootInteractive plays a central role in expert projects such as digital signal processing for Run 3 \cite{ALICETPCCommonMode:2023ojd}, optimization and validation of track reconstruction for Run 2 and Run 3 \cite{IvanovTracking_2022}, MC/data mapping, TPC data volume studies, and differential quality assurance and quality control. It has also been used in the development of the particle identification algorithms and magnetic monopole reconstruction studies (in collaboration with the DUNE experiment) for high dE/dx (mean energy loss per distance traveled), low momentum, and spallation product tracking.
 In summary, RootInteractive stands at the forefront of advanced data analysis in the challenging scientific context of ALICE CERN.

%\section{RootInteractive}
%\label{RootInteractive}
\begin{figure*}
\centering
%\vspace*{5cm}       % Give the correct figure height in cm
\includegraphics[width=\textwidth,clip]{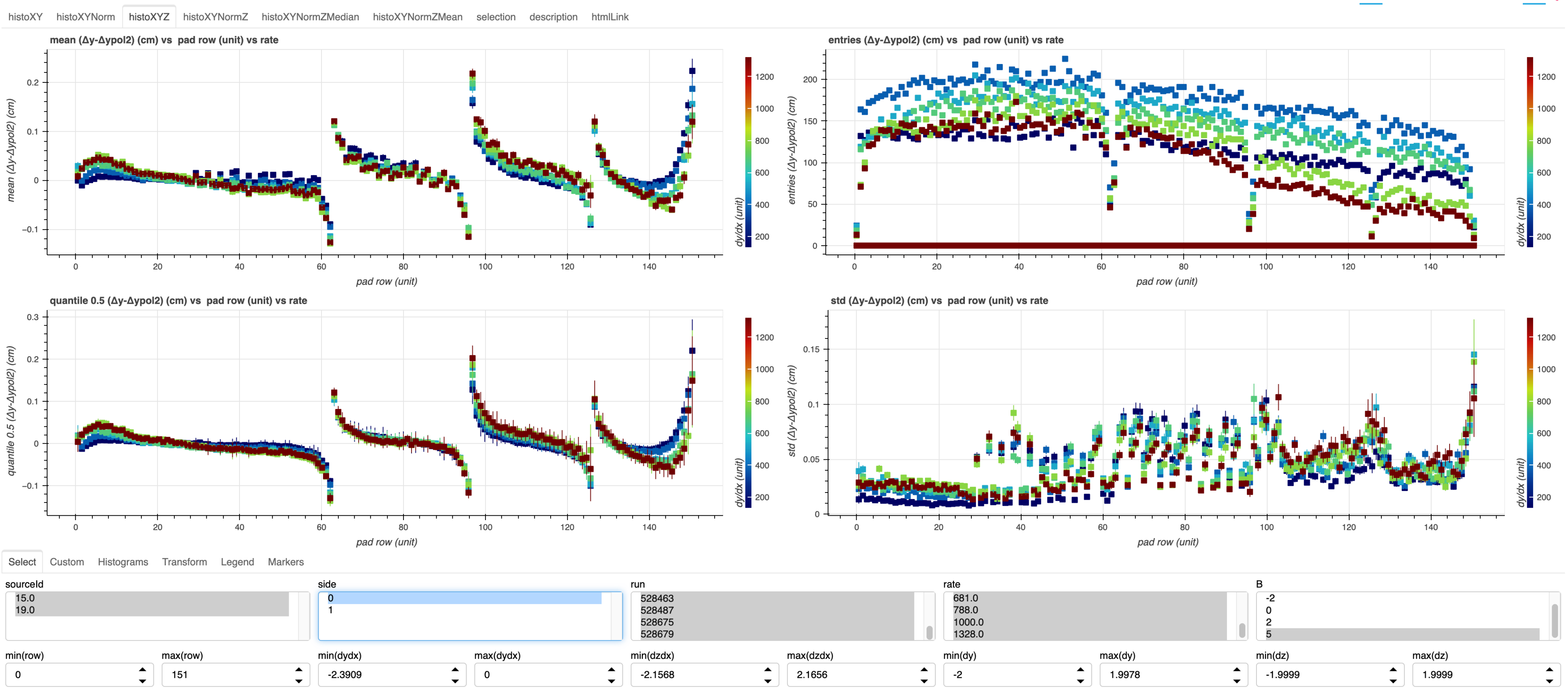}
\caption{
Example of an interactive dashboard created with the getDefaultVars function of the RootInteractive template. The goal of this dashboard is to compare the ALICE TPC space point distortion estimator based on two factors: position in the detector (pad row) and rate (represented by the colour axis).
The layout of the dashboard is very versatile. It offers views for 1D, 2D and 3D aggregation, both with and without data normalisation. These different views can be accessed via tabs so that the aggregated data and the different iterations of the machine learning predictions can be easily compared.
On the snapshot you can see the 2D layout. It shows important statistical quantities such as mean, median and root mean square (rms) for a given selection (e.g. side, magnetic field, position, etc.).
}
\label{fig1}       % Give a unique label
\end{figure*}

The framework presented here, RootInteractive, serves as a versatile tool for interactive statistical aggregation and visualisation of multidimensional data, compatible with both ROOT\cite{aamodt2008alice} and native Python data frame formats such as Pandas \cite{pandas} and Modin \cite{modin}.
The code provides extensive support for various ROOT data structures and classes, including TTree, TTreeFormula, Aliases, TFormula and static Root/AliRoot functions. Work is also being done to simplify compatibility with RDataFrame \cite{piparo2019rdataframe,guiraud2023rdataframe} and awkward (PyHep) arrays \cite{awkward2023}.
One can use this framework with various data sources, including PyRoot (AliRoot/O2) data structures. Importantly, it works seamlessly with pandas/modin alone, so one does not need to install the ROOT package. Internally, these data structures are converted into either Bokeh \cite{bokeh} CDS (ColumnDataSource) for simple scatter visualization or our own RootInteractive CDS extension, which allows for a variety of operations with N-dimensional histograms, projections, and aggregated data.

The content of RootInteractive includes the following main aspects:

\begin{itemize}
    \item It provides an interactive and highly customisable visualisation solution for both non-binned and binned data.
    \item It allows interactive operations such as n-dimensional histograms, projections and the extraction of derived aggregated information.
\item The application can be used in different configurations, e.g. as a client/server application integrated with Jupyter and Bokeh.
 \item Users can use the Bokeh standalone dashboard independently as a client application, without having to install software or require a stable internet connection to a server. This is the most common way to use RootInteractive. We can include interactive dashboards as a valuable resource in meeting agendas, for example.
    \item The system supports both lossy and lossless data compression, enabling efficient data transfer from server to client. In typical use cases, a factor ~10 is achieved.
    \item RootInteractive interfaces seamlessly with ROOT - RDataFrame -awkward - Pandas/Modin tools.
    \item Work is underway to further simplify RDataFrame's C syntax by using Domain-specific language for slicing and joins (inspired by Python syntax and Pandas/Modin joins). The Python-like syntax is translated into the C++ template functions and used as JIT or in C macros.
\end{itemize}

This code empowers interactive visualization, histogramming, and data aggregation in N-dimensions on the client side, facilitating advanced data analysis tasks.

\section{Exploring Symmetries, Alarms, and Invariants in Multi-Dimensional Data Analysis}
\label{ExploringSymmetries}
In our project, we focus on the use of symmetries and invariants (within uncertainties) for multidimensional data analysis. We optimize the handling of normalized data that includes data-analytical model, MC-real data,  data-symmetry, data-reference data, and data-machine learning prediction across different dimensions.
Data normalization leads to reduced RMS scatter and the ability to implement alarms and outlier detection based on statistical significance, e.g. identifying cases where (data model) exceeds N$\sigma$ or using likelihood-based methods.
These methods are actively applied in various ALICE projects and include considerations of temporal invariance (e.g. referencing the data to an average run), spatial invariance (taking into account rotational and mirror symmetry), magnetic field symmetry, comparing the data with analytical models, evaluating different machine learning models and assessing the smoothness of the data.

As an ALICE application example of symmetry in data and ML models, we train models where we assume symmetry and compare them to a regression or statistical aggregation without the symmetry assumption. We do this by including or not including the variable with the expected symmetry in the Machine learning regression or aggregation. 

For example, in the case of space point distortion, the particle production is $\varphi$-symmetric, so the space charge density in the TPC detector is also $\varphi$-symmetric, and accordingly the E-vectors of the distortion should also have the same symmetry. A deviation from symmetry either indicates a problem, or the symmetry is broken by an additional effect that we have to take into account (e.g. a non-$\varphi$-symmetric conversion factor from ionisation to space charge) in correction respectively in MC simulation. After correcting (normalising over the functional composition) for the known effects, the symmetry should be restored and data consistency can be assessed.

\section{RootInteracive - Machine learning, ML validation and data aggregation}
\label{machineLearnigAnfAggregation}
RootInteractive integrates external machine learning models and provides wrappers for models such as RandomForest \cite{scikit-learn} and Extreme Gradient BDT \cite{Chen:2016:XST:2939672.2939785xgboost} to estimate local parameters of the PDF function (which are used as estimators for the reducible and irreducible error machine learning model).
For example, in optimizing the TPC space point distortion model developed by ALICE, we used RootInteractive to compare an external U-net model \cite{ronneberger2015unet} with a simpler data-driven approach using RandomForest. This involved optimizing the parameters of the models and the cost functions.

RootInteractive performs interactive visualization, data aggregation and invariance validation on the client side, processing significant amount of unbinned data, typically between O($10^6$) and O($10^8$) elements (rows x attributes). The upper limit depends on memory and data transfer and is typically O(1GB) on the client side. To access even larger amount of data, two main approaches are used: domain-specific sampling (e.g. in ALICE, it provides more uniform momentum or particle type distribution) and pre-aggregation on the server. This pre-aggregated or skimmed information can later be reweighted in subsequent RootInteractive sessions on the client.
Pre-aggregated data sources typically include local statistics such as mean, median, count and standard deviation obtained from unbinned predictions (from Machine learning regression). Alternatively, we aggregate data using local kernel regression parameters tabulated on a regular mesh. In RootInteractive we use C (HistoND) and Python libraries like Pandas \cite{pandas} and Modin \cite{modin} for multidimensional operations like 'group by' and rolling statistics.

\section{RootInteractive statistics and Machine learning wrappers}
\label{RootInteractive statistics}
RootInteractive provides simple local statistics on the regular grid, including mean, median, RMS, and quantiles, for model validation on the client side (see example snapshot \ref{fig1}).

We also introduce new features:
\begin{itemize}
    \item Generalized kernel linear regression on the client, similar to that on the server. This uses multi-dimensional group-by, rolling statistics, and local kernel fitting to represent smooth functions.
    \item Ongoing development of client-side ML prediction using WebAssembly (wasm) \cite{WebAssemblyCoreSpecification2} and ONNX \cite{onnxruntime}. This allows us to parameterize machine learning models and define parameterized derived variables. For example, we can compute derivatives of machine learning predictions, perform systematic error studies on the client side, and perform numerical derivatives with varying input parameters of machine learning models.
    \item A local linear forest, which is a local linear regression with a kernel defined by a random forest, was originally introduced in GRF (Generalised Random Forest) \cite{athey2019generalized}. However, the original implementation is very computationally intensive when it comes to predictions. To mitigate this, we are working on a cached version that incorporates local derivatives in the nodes, similar to the approach used in the previous ALICE software framework \cite{AliRoot}) framework in the AliNDLocalRegression class on the fixed grid. This optimisation makes the prediction process more efficient.
\end{itemize}

\section{Interactive visualization, histogramming, and data aggregation in N-dimensions on client}
\label{InteractiveVisualization}

Interactive visualisation, histogram generation and N-dimensional data aggregation on the client side are controlled by a series of Python dictionaries and arrays. These declarations serve as inputs to the bokehDrawSA function, which can be used to create a variety of graphical elements such as scatter plots and N-dimensional histograms, as well as projection statistics, whether they are binned or unbinned. Essentially, these declarations define the data sources for bokeh and allow the development of derived variables and aggregated statistics to enrich the client-side visualization.
bokehDrawSA uses declarative programming, an approach that allows developers to express computational logic without having to explicitly script every step of the process. This methodology simplifies the programming effort because developers only need to describe the desired program results, rather than specifying in detail each command or step to achieve those results.
In practice, the configuration of the interactive visualization relies on six arrays/dictionaries, as shown below:

\begin{python}
bokehDrawSA.fromArray(df,selection,figureArray,widgetParams, 
layout=figureLayoutDesc, tooltips=tooltips, 
parameterArray=parameterArray,
widgetLayout=widgetLayoutDesc, sizing_mode="scale_width", 
nPointRender=300,aliasArray=aliasArray, 
histogramArray=histoArray,arrayCompression=arrayCompression)
\end{python}

These arrays, in particular figureArray, histogramArray, aliasArray, layout, widgetLayout, and parameterArray, collectively contribute to the construction and creation of interactive visualizations and provide a streamlined approach to the complex display and analysis of data. The parameter names used in the declaration directories conform to the bokeh naming convention for graphical elements. For histogramming and statistical aggregation, we have taken inspiration from Numpy and Pandas.

To simplify the creation of interactive visualizations, we use "predefined template functions" that provide predefined configurations (dictionaries) that can later be extended in the user's Python code. Our main goal is to simplify the process of automatic multidimensional differential validation, and to this end, we have developed several template functions with parameterisable differences of function, reference function and scale function. 

One of these functions, getDefaultVars, is tailored to create default variables suitable for multidimensional data and automatic user-defined normalization. It provides configurations such as aliasArray, variables, parameterArray, widgetParams, widgetLayoutDesc, histoArray, figureArray and figureLayoutDesc.

\begin{python}
def getDefaultVars(normalization=None, variables=None, defaultVariables={}, weights=None, multiAxis=None)
\end{python}

\begin{figure*}
\centering
%\vspace*{5cm}       % Give the correct figure height in cm
\includegraphics[width=\textwidth,clip]{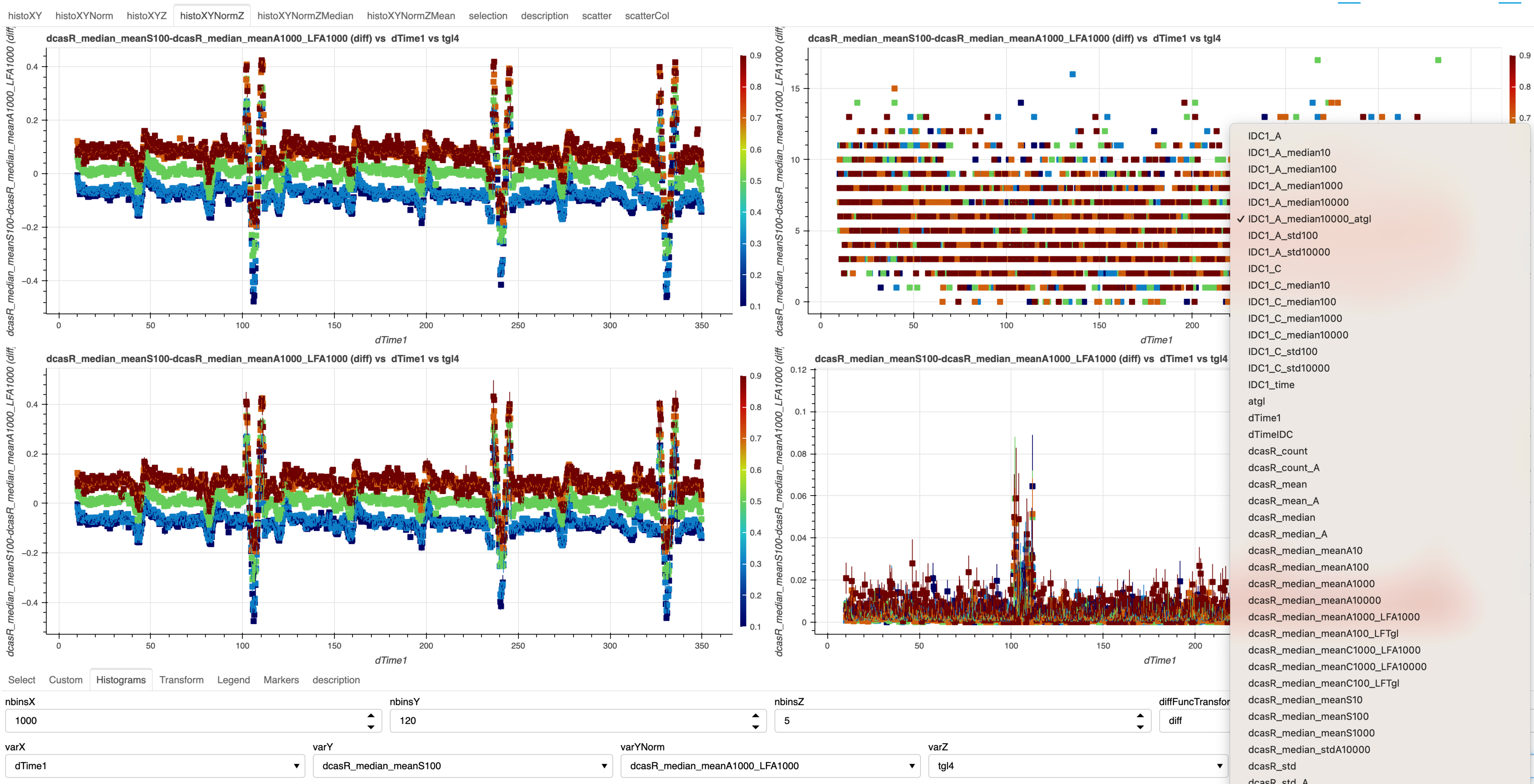}
\caption{
A snapshot of a real use-case l interactive dashboard within ALICE, generated using getDefaultVars. This dashboard is used for comparing ALICE TPC \cite{alme2010aliceTPC} currents (IDC) and TPC tracking bias (<DCA>) time series. The layout includes functions for 1D, 2D, and 3D aggregation, facilitating comparisons between aggregated data and various iterations of machine learning predictions. Additionally, it provides local aggregated statistics. You can find a partial list of the input variables for interactive aggregation on the right side of the snapshot (expanding select for the varZ selection)
}
\label{figSnapshot2}       % Give a unique label
\end{figure*}

With this function, we create a predefined layout containing 1D, 2D and 3D aggregations (a user extension must create an n-dimensional selection layout). The parameters for aggregation and normalization are based on the lists of input variables. The client-side normalization process provides flexibility and supports different methods such as "delta," "ratio", "log ratio" and pulls ($\Delta/\sigma$). In example snapshot fig. \ref{figSnapshot2}, the diff function is selected using multiselect widget - difFunction.

In example above the following code for aliasArray  and parameterArray was generated:
\begin{itemize}
    \item Generated aliasArray:
\end{itemize}
\begin{python}
  [{'fields': ['varY', 'varYNorm'],
  'name': 'diffFunc',
  'parameters': ['diffFuncTransform'],
  'v_func':  """
       if($output == null || $output.length !== varY.length){ 
           $output = new Float64Array(varY.length) 
       } 
        if(diffFuncTransform=='diff'){
           for(let i=0; i<$output.length; i++){ 
               $output[i] = varY[i]-varYNorm[i] 
           } 
       } 
        else if(diffFuncTransform=='ratio'){
           for(let i=0; i<$output.length; i++){ 
               $output[i] = varY[i]/varYNorm[i] 
           }                 
       } 
        else if(diffFuncTransform=='logRatio'){
           for(let i=0; i<$output.length; i++){ 
               $output[i] = Math.log(varY[i])/Math.log(varYNorm[i]) 
           }                 
       } 
       return $output 
    """ }]
\end{python}
\begin{itemize}
    \item Generated parameterArray subset:
\end{itemize}
\begin{python}
    {'name': 'diffFuncTransform', 'value': 'diff', 'options': ['diff', 'ratio', 'logRatio']},
    {'name': 'varY', 'value': 'dLX_neg_3000_mean_G0', 'options': ['dLX_neg_3000_mean_G0', 'dLX_neg_3000_mean_G1', ....
\end{python}

\section{Enhancing Interactive Analysis and Code Efficiency through Representative Down-Sampling}

To enable interactive physics analysis, optimize code performance, and generate training datasets for machine learning techniques, representative down-sampling is employed. This process ensures a roughly flat distribution in variables of interest. In our specific physical use case, we utilize a combination of minimum bias data and down-sampled data with a roughly flat distribution in particle momenta, event multiplicity, and PID. Depending on the dataset, our typical target value is \(O(10^{-2}-10^{-4})\). This down-sampling process occurs outside of RootInteractive.

To achieve the desired roughly flat distribution, we leverage several effective parameterizations with significant precision. Rather than requiring an exact distribution, weights are typically stored for subsequent re-weighting in the RootInteractive analysis. The minimum bias sample serves as a control mechanism for this methodology.

\section{Conclusions}

RootInteractive was developed as a tool that simplifies multidimensional data analysis and allows us to effectively process data in all relevant dimensions. This tool aims to fit and visualise multidimensional functions, including uncertainties and biases, validate assumptions and approximations, facilitate the composition of parametric and non-parametric functions, and use symmetries to define multidimensional "invariant" functions and alarms.

RootInteractive offers functions for interactive visualization of both unbinned and binned data, n-dimensional histogramming and projection, and extracting aggregate information, both on the server (Python/C++) and client (JavaScript) in browser. By employing a combination of lossy and lossless data compression techniques, RootInteractive allows interactive analysis of datasets containing millions of entries and dozens of attributes within a web browser, typically consuming around 0.5-1 GB of memory. 
Through representative downsampling (typically 1-0.1\% of data) followed by reweighting or pre-aggregation on the server or batch farm, it enables interactive multidimensional analysis of ALICE's extensive monthly/annual statistics for calibration, reconstruction validation, quality assurance, quality control, and statistical/physical analysis.

Thanks to its versatility and ease of use, RootInteractive plays a central role in ALICE expert projects such as digital signal processing for Run3, optimisation and validation of reconstruction for Run2 and Run3, MC /data mapping, TPC data volume studies and differential quality assurance and quality control. It has also been used in the development of the particle identification algorithm and in magnetic monopole reconstruction studies (in collaboration with the DUNE experiment) for high dE/dx (mean energy loss per distance traveled), low momentum and spallation product tracking.

\label{Conclusion}

\section{Future Work}

% Your future work content goes here
% You can reference or rephrase the moved text here

Future work includes continuing efforts to streamline the C syntax of RDataFrame.
This endeavor includes the application of a domain-specific language for tasks such as slicing, joins and rolling statistics that are inspired by the syntax of Python and the joining methods used in Pandas/Modin. The planned approach involves the development of a Python-like syntax that is translated into C++ template functions and used either for just-in-time (JIT) compilation or for integration into C macros.
As explained in the previous sections, there are also extensive ongoing developments focussing on the integration of predictions for machine learning on the client side through the integration of
WebAssembly \cite{WebAssemblyCoreSpecification2} and ONNX \cite{onnxruntime}.

% ... (remaining sections and document structure)

%
% BibTeX or Biber users please use (the style is already called in the class, ensure that the "woc.bst" style is in your local directory)
% \bibliography{name or your bibliography database}

\bibliography{chepMI2023}

%\begin{thebibliography}
%\bibitem{Abelev:2014ffa}
%  B.~B.~Abelev {\it et al.} [ALICE Collaboration],
%  %``Performance of the ALICE Experiment at the CERN LHC,''
%  Int.\ J.\ Mod.\ Phys.\ A {\bf 29} (2014) 1430044.
%\end{thebibliography}

\end{document}